\documentclass[article]{rmaa}

\usepackage{paralist}

\usepackage{psfrag,color}

\title{A Hydrodynamical Mechanism for Generating Astrophysical Jets}

\author{
  Xavier Hernandez,\altaffilmark{1}
  Pablo. L. Rend\'on\altaffilmark{2}
  Rosa G. Rodr\'{\i}guez-Mota\altaffilmark{2}
  A. Capella\altaffilmark{3}
}
\altaffiltext{1}{Instituto de Astronom\'\i{}a, UNAM, M\'exico.}

\shortauthor{Hernandez et al.}
\shorttitle{A Hydrodynamical Mechanism for Generating Astrophysical Jets}

\fulladdresses{
\item X. Hernandez: Instituto de Astronom\'\i{}a, Universidad Nacional
  Aut\'onoma de M\'exico, Apartado Postal 70264, CD Universitaria, CP 04510
  , M\'exico DF, M\'exico (\email{xavier@astro.unam.mx})
\item Pablo. L. Rend\'on: Centro de Ciencias Aplicadas y Desarrollo Tecnol\'ogico, 
Universidad Nacional Aut\'onoma de M\'exico, AP 70--186, M\'exico D.F. 04510, M\'exico
\item Rosa G. Rodr\'{\i}guez-Mota: Centro de Ciencias Aplicadas y Desarrollo Tecnol\'ogico, 
Universidad Nacional Aut\'onoma de M\'exico, AP 70--186, M\'exico D.F. 04510, M\'exico
\item A. Capella:Instituto de Matematicas, Universidad Nacional Aut\'onoma de M\'exico, M\'exico D.F. 04510, M\'exico. 
}

\listofauthors{Hernandez, Rend\'on, Rodr\'{\i}guez-Mota \& Capella}
\indexauthor{Hernandez, X.}
\indexauthor{Rend\'on, P.L.}
\indexauthor{Rodr\'{\i}guez-Mota, R.G.}
\indexauthor{Capella, A.}

\abstract{Whenever in a classical accretion disk the thin
disk approximation fails interior to a certain radius, a transition from Keplerian to radial infalling trajectories
should occur. We show that this transition is actually expected to occur interior to a certain critical radius,
provided surface density profiles are steeper than $\Sigma(R) \propto R^{-1/2}$, and further, that it probably corresponds to
the observationally inferred phenomena of thick hot walls internally limiting the extent of many stellar accretion disks. 
Infalling trajectories will lead to the 
convergent focusing and concentration of matter towards the very central regions, most of which will simply be swallowed 
by the central object. We show through a perturbative hydrodynamical analysis, 
that this will naturally develop a well collimated pair of polar jets. A first
analytic treatment of the problem described is given, proving the feasibility of purely hydrodynamical mechanisms 
of astrophysical jet generation. The purely hydrodynamic jet generation mechanism explored here complements existing
ideas focusing on the role of magnetic fields, which probably account for the large-scale collimation and
stability of jets.}

\resumen{Whenever in a classical accretion disk the thin
disk approximation fails interior to a certain radius, a transition from Keplerian to radial infalling trajectories
should occur. We show that this transition is actually expected to occur interior to a certain critical radius,
provided surface density profiles are steeper than $\Sigma(R) \propto R^{-1/2}$, and further, that it probably corresponds to
the observationally inferred phenomena of thick hot walls internally limiting the extent of many stellar accretion disks. 
Infalling trajectories will lead to the 
convergent focusing and concentration of matter towards the very central regions, most of which will simply be swallowed 
by the central object. We show through a perturbative hydrodynamical analysis, 
that this will naturally develop a well collimated pair of polar jets. A first
analytic treatment of the problem described is given, proving the feasibility of purely hydrodynamical mechanisms 
of astrophysical jet generation. The purely hydrodynamic jet generation mechanism explored here complements existing
ideas focusing on the role of magnetic fields, which probably account for the large-scale collimation and
stability of jets.}

\addkeyword{accretion, accretion discs}
\addkeyword{hydrodynamics}
\addkeyword{galaxies:jets}
\addkeyword{ISM:jets and outflows}

\begin{document}
\maketitle

\section{Introduction}

Astrophysical jets occur over a large range of astronomical scales, from the stellar Newtonian scales of HH objects 
(e.g. Reipurth \& Bally 2001), to the relativistic cases of microquasars and gamma ray bursts (e.g. Mirabel \& Rodriguez 1999,
M\'esz\'aros 2002), to the Mega-parsec extragalactic scales of AGN jets (e.g. Marscher et al. 2002). Although the processes of
propagation and collimation appear to be relatively well understood in terms of the interplay between the hydrodynamics of the 
problem (Scheuer 1997) and magnetic fields (e.g. Blandford 1990), the precise mechanism of jet generation remains as the
most uncertain part of the system.   

Existing jet generation mechanisms focus on the interaction between the rotating material in the inner regions of the accretion disk
and the magnetic field of the central (rotating or otherwise) star, or perhaps an external magnetic field threading the disk.
See for example Blandford \& Payne (1982),  Henriksen \& Valls-Gabaud (1994), Meier et al. (2001), Price et al. (2003). 
Although generically successful, detailed comparisons with observations do dot always yield consistent 
results, e.g. Ferreira et al. (2006). Further, the details of relative orientation of stellar spin, accretion disk and magnetic 
field configurations, have to be supplied in somewhat highly specific manners.  

As an alternative, we explore the possibility of hydrodynamical jet generation mechanism where magnetic fields play no r\^ole,
but only the intrinsic hydrodynamical physics of the interplay between the accretion and the central star wholly determine the 
characteristics of the jet. The main obstacle to such a scheme is the presence of a centrifugal barrier associated with the angular 
momentum content of the material in the accretion disk. We show however, that the thin disk approximation for accretion disks, which 
is equivalent to the assumption of quasi-circular orbits for the disk (e.g. Pringle 1981), is expected to break down internal to a 
certain critical radius, resulting in a transition to quasi-radial flow for the disk. Then, an analytic first order
perturbation treatment serves to demonstrate the feasibility of purely hydrodynamical jets. We show also that many generic features of 
observed astrophysical jets across a wide range of scales, can be naturally accounted for in the general model presented.
The presence of magnetic fields in jet phenomena is evident empirically, however, it is not impossible that their main contribution 
to the problem could be restricted to the radiation of the jet material and to its long range collimation, with purely hydrodynamical 
physics playing a part in the actual jet generation mechanisms.

The problem of jet formation has been studied extensively in the context of classical hydrodynamics, most often
regarding fluid-body interactions. The appearance of stable coaxial jets resulting from radially-symmetric
velocity fields over thin fluid sheets has been established by, among others, Taylor (1960). The r\^ole
played by both the magnitude and the direction of velocity in the formation of this type of jet is the
subject of a theoretical study by Glauert (1956), where it is shown that at the point at which the jet
forms, a large velocity gradient is observed and momentum flux is constant, with horizontal momentum being
transformed into vertical momentum. Similar results were obtained by King \& Needham (1994), who provide an
asymptotic study of a jet formed by a vertical plate accelerating into a semi-infinite expanse of
stationary fluid of finite depth with a free surface and a gravitational restoring force. It is found that
as the fluid approaches the plate in a horizontal direction, a gradual rise in free-surface elevation
occurs. Eventually, a thin region is reached where vertical velocity dominates horizontal velocity as a
consequence of the fluid finding it more difficult to overcome the inertia it would meet by continuing its
horizontal motion than escaping vertically towards the low-pressure free surface. In essence, this same
mechanism can allow for jet formation in the context of the problem we study here.

In section 2 we develop the criterion for transition to radial flow in a standard accretion disk.
The perturbative solution to the resulting problem of a radially infalling disk is then developed in
section 3. Section 4 presents trajectories for particular cases of the solution obtained in
the previous section, in  dimensionless units. Finally, we give our conclusions in section 5.

\section{The transition to Radial Flow}

We start from a standard thin accretion disk where material orbits on quasi-circular Keplerian
orbits around a central star of mass M. Assuming axial symmetry and cylindrical coordinates, the total angular 
momentum of a ring at radius $R$ of width $\Delta R$ will be: 

\begin{equation}
L=2 \pi R^{3} \Delta R \Sigma(R) \Omega(R)
\end{equation}
where $\Sigma$ and $\Omega$ are the surface density and orbital frequency profiles of the disk, respectively.
If the breaking torque on a given ring of the disk, due to its exterior ring is denoted by $\tau(R)$,
the total breaking torque on the ring at radius $R$ will be:

\begin{equation}
\tau_{B} = \tau(R+ \Delta R) - \tau(R).
\end{equation}

The rate of change of the angular momentum of the ring in question can now be calculated as:

\begin{equation}
\dot{L}=~\tau_{B}=~\frac{d\tau}{dR} \Delta R.
\end{equation}

If at any radius a substantial fraction of the angular momentum of the ring is lost through viscous torques
over an orbital period, the assumption of quasi-circular orbits will break down, and the disk will make a 
transition to mostly radial orbits. This condition can be stated as:

\begin{equation}
\frac{L}{2 \dot{L}}< \frac{2 \pi}{\Omega}.
\end{equation}

Substitution of equations (1) and (2) into the above yields:

\begin{equation}
R^{3} \Omega^{2} \Sigma < 2 \left( \frac{d \tau}{dR} \right),
\end{equation}
as the condition for the onset of radial flow in the disk. 

We can now introduce a model for the torques in terms of the rate of shear and the effective viscosity $\nu$
(e.g. Von Wiezs\"{a}cker 1943, Pringle 1981) as:

\begin{equation}
\tau=2 \pi R^{3} \nu \Sigma \frac{d\Omega}{dR}.
\end{equation}

Taking $\Omega^{2} = GM/R^{3}$ to substitute the above equation into condition (5) gives

\begin{equation}
(GM)^{1/2} \Sigma < -6 \pi \frac{d}{dR} \left(R^{1/2} \nu \Sigma \right).
\end{equation} 

To proceed further we can take for example, $\nu=$constant and a model for the disk surface density
profile of the form:

\begin{equation}
\Sigma=\Sigma_{0} \left(\frac{R}{R_{0}} \right)^{-n},
\end{equation}
of the type often used in models of accretion disks when fitted to observations 
(e.g. Hughes et al. 2009). Use of the two above forms for $\nu$ and $\Sigma(R)$ reduces condition(7) to:

\begin{equation}
6 \pi \nu (n-1/2) > (G M R)^{1/2}.
\end{equation}

The above condition will always be met in any accretion disk with $n>1/2$, interior to a transition radius
$R_{T}$ given by:

\begin{equation}
R_{T} = \frac{[6 \pi \nu (n-1/2)]^{2}}{GM}.
\end{equation}

It is interesting that directly observed accretion disks have spectra which when modelled typically yield
$n \sim 1$, in general $0<n<1.5$, e.g. Hartmann et al. (1998), Lada (2006), Hughes et al. (2009). 
This leads one to expect the transition to radial flow to occur in many of
the stellar accretion disks, interior to radii given by equation (10).

Finally, we can explore the consequences of
introducing an $\alpha$ prescription (Shakura \& Sunyaev 1973) into condition (9), $\nu=\alpha c H$, where
$c$ and $H$ are respectively the sound speed and height of the disk, with $\alpha$ a dimensionless number, 
yielding the dimensionless condition:

\begin{equation}
 {\mathcal M}  \left( \frac{R}{H} \right) < 6 \pi \alpha (n-1/2),
\end{equation}
where  ${\mathcal M}$ is the ratio of the Keplerian orbital velocity in the disk to the sound speed in
the disk. Although the above condition is only valid for constant $\nu$, it illustrates the equivalence
between the assumption of quasi-circular orbits for the material in the disk, and the assumption that the disk
is thin. We see that the breakdown of the assumption of quasi-circular orbits (condition 9), corresponds
to the point where disk becomes fat.

Writing ${\mathcal M}$ in astrophysical units as:

\begin{equation}
 {\mathcal M} = 1.8 \left( \frac{200 K}{T} \right)^{1/2}  \left( \frac{50 AU}{R} \right)^{1/2}  
\left( \frac{M}{0.5 M_{\odot}} \right)^{1/2},
\end{equation}

\noindent we see that for values typical of what is observationally inferred for the stellar mass and position and 
temperature of accretion disk walls in T Tauri protoplanetary disks, those indicated in the above equation 
(e.g. D'Alessio et al. 2005, Espaillat et al. 2007, Hughes et al 2009), one approaches the breakdown of the thin 
disk approximation, ${\mathcal M}<1$, and consequently the transition to radial flow.

Taking typical inferred values for $R/H$ at the wall of $\sim 5$ (e.g. D'Alessio et al. 2005, Espaillat et al. 2007, 
Hughes et al 2009), we can now write the condition for the transition to radial flow, locally at the wall, as:

\begin{equation}
10=6 \pi \alpha (n-1/2).
\end{equation}

We see that for a standard value of $n\sim1$ the above equation implies a reasonable value of $\alpha \sim 1$
at the thick wall, significantly higher than the values of $\sim 0.01$ and lower, which apply for the body of the 
thin accretion disk beyond this radius. A substantial increase of $\alpha$ as $R/H$ decreases is expected in any turbulence
driven viscosity model for accretion disks, e.g. Firmani, Hernandez \& Gallagher (1996), in the context of galactic disks.

In terms of the debate surrounding the inference of inner holes in observed accretion disks, many solutions have been 
proposed in terms of disk clearing mechanisms; grain growth (e.g. Strom et al. 1989, Dullemond \& Dominik 2005), 
photoevaporation (e.g. Clarke et al. 2001), magnetorotational instability inside-out clearing (e.g. Chiang \& Murray-Clay 2007,
Dutrey et al. 2008), binarity (e.g. Ireland \& Kraus 2008) and planet-disk interactions (e.g. Rice et al. 2003). 
None of the above is entirely satisfactory, as noted by Hughes et al. (2009), mostly due to their incompatibility 
with a steady state solution. An alternative solution under the proposed scenario, is that there is no actual disk material clearing, 
only a transition to predominantly radial flow at the thick wall, and consequently a shear-free flow interior to this point. Once the 
disk heating mechanism is removed, as one should expect from the analysis presented in this section, the inner disk disappears from sight.

This transition at a thickened inner boundary is qualitatively what ADAF models propose. In those models, substantial turbulent 
and thermal pressure remain within the inner flow. Still, given the present lack of a definitive jet generation mechanism, 
we consider it interesting to explore the consequences of a model where cooling drives 
the flow to a $\nabla \vec{P_{0}}/ \rho << \nabla \Phi$ condition. As we show in the following sections, we prove that such a 
situation will quite naturally yield purely hydrodynamical jet solutions, which we find encouraging.

\section{Hydrodynamical Jet Solutions}
\label{hydro}

We shall now model the physical situation resulting from the scenario described above as an axially
symmetrical distribution of gas in free fall towards a central star of mass $M$. Taking a spherical
coordinate system with $\theta$ the angle between the positive vertical direction and the position vector $\vec{r}$
we have:

\begin{equation}
\frac{1}{r^{2}} \frac{\partial(r^{2} \rho V)}{\partial r} + \frac{1}{r sin(\theta)}\frac{\partial(sin(\theta) \rho U)}
{\partial \theta}=0,
\end{equation}

\begin{equation}
V\frac{\partial V}{\partial r} +\frac{U}{r}\frac{\partial V}{\partial \theta} -\frac{U^{2}}{r}=
-\frac{1}{\rho}\frac{\partial P}{\partial r} - \frac{G M}{r^{2}},
\end{equation}

\begin{equation}
V\frac{\partial U}{\partial r} +\frac{U}{r}\frac{\partial U}{\partial \theta} +\frac{V U}{r}=
-\frac{1}{r \rho}\frac{\partial P}{\partial \theta},
\end{equation}

\noindent for the continuity equation, and the radial and angular components of Euler's equation. In the above, $V$, $U$, $\rho$
and $P$ give the radial velocity, angular velocity, matter density and pressure, respectively. We have neglected temporal
derivatives, as we are interested at this point, in the characteristics of steady state solutions, although the temporal derivatives are reinstated in the Appendix for the purpose of stability analysis. We take as a background
state a free-falling axially symmetrical distribution of gas described by $V_{0}=-(2GM/r)^{1/2}$, $U_{0}=0$
and all components of $\nabla \vec{P_{0}}/ \rho_{0}$ negligible, a consistent solution to eqs.(15) and (16), used 
only in the description of this background. Having ignored the inclusion of the radius at which
the transition to radial flow takes place in the choice of $V_{0}$ limits the validity of the analysis to radial scales
along the plane of the disk much smaller than $R_{T}$. This is justified by the fact that jets appear as phenomena 
extremely localised towards $R \rightarrow 0$. We now take a density profile given by:

\begin{equation}
\rho_{0}(r,\theta)=f(r)g(\theta),
\end{equation}

\noindent were $f(r)$ is a dimensionless function of $r$ and $g(\theta)$ describes the polar angle dependence of the infalling
material, for example, one can ask for $g(\theta=\pi/2)=\bar{\rho}_{0}$, diminishing symmetrically towards the poles. 
The choice of this last function will determine the details of the problem, and can be thought of as something of the type

\begin{equation}
g(\theta)=\bar{\rho}_{0}e^{-\left( \frac{\theta- \pi/2}{\sqrt{2} \theta_{0}}   \right)^{2}}
\end{equation}

\noindent with $\bar{\rho}_{0}$ a normalisation constant and $\theta_{0}$ a form constant describing the flattening of the
disk of infalling material. However, we shall mostly leave results indicated in terms of $g(\theta)$. The continuity 
equation (14) now fixes $f(r)$ through:

\begin{equation}
-\frac{g(\theta)}{r^{2}} \frac{d}{d r} \left[ (2 G M)^{1/2} r^{3/2} f(r) \right]=0,
\end{equation}

\noindent and hence $f(r) r^{3/2} =cte.$, which completes the description of the background state through:

\begin{equation}
\rho_{0}(r,\theta)=\left( \frac{\bar{r}}{r} \right)^{3/2} g(\theta),
\end{equation}

\noindent with $\bar{r}$ a constant which determines the point at which $g(\theta=\pi/2)$ $\Rightarrow$ $\rho_{0}=\bar{\rho}_{0}$.

We now study the first order departures from the assumed background state through a perturbative approach, to explore
the possibility that such departures might naturally give rise to a hydrodynamical jet.
Regarding these first order perturbations, we shall also be interested in steady state solutions given by:

\begin{equation}
V(r,\theta)=V_{0}(r) + V_{1}(r, \theta),
\end{equation}

\begin{equation}
U(r,\theta)=0 + U_{1}(r,\theta),
\end{equation}

\begin{equation}
\rho(r,\theta) = \rho_{0}(r,\theta) +\rho_{1}(r,\theta),
\end{equation}

\noindent where quantities with subscript (1) denote the perturbation on the background solution.
Notice that $V_{0}$ is, as required by the assumed background state, a function only of the radial coordinate $r$.
For the background state assumed, the perturbations on it will be solved for in a fully self consistent manner. 
Writing eqs.(14), (15) and (16) to first order in the perturbation one obtains after rearranging terms:

\begin{equation}
g(\theta) \frac{\partial \left(r^{1/2} V_{1}\right)}{\partial r} - 
B\frac{\partial \left( r^{3/2} \rho_{1} \right)}{\partial r}=
\frac{-1}{r^{1/2} sin(\theta)} \frac{\partial \left( sin(\theta) g(\theta) U_{1} \right) }{\partial \theta},
\end{equation}

\begin{equation}
\frac{\partial \left( V_{1}/r^{1/2} \right) }{\partial r} = \frac{A r^{3/2}}{g(\theta)} \frac{\partial \rho_{1}}{\partial r},
\end{equation}

\begin{equation}
\frac{\partial \left( r U_{1}  \right)}{\partial r} = \frac{A r^{2}}{g(\theta)} \frac{\partial \rho_{1}}{\partial \theta},
\end{equation}

\noindent In the above we have assumed an isothermal equation of state for the perturbation $P_{1}=c^{2}\rho_{1}$, an assumption
often used in the modeling of astrophysical jets, e.g. the T Tauri jets observed and modelled by Hartigan et al. (2004). This 
idealised case serves to illustrate clearly the consequences of the physical setup being considered, as it 
allows for an analytic solution. The generalisation to more realistic adiabatic, polytropic or otherwise equations of 
state can be performed numerically, and can be expected to yield qualitatively similar results, although interesting differences 
in the details can be expected to emerge, which will be considered latter. In the above three equations we have introduced
the constants $A=(c^{4}/2GM\bar{r}^{3})^{1/2}$ and $B=(2GM/\bar{r}^{3})^{1/2}$.

To make further progress we can attempt a solution through the method of separation of variables, proposing a solution
of the form: $V_{1}=V_{r} V_{\theta}$, $U_{1}=U_{r} U_{\theta}$, $\rho_{1}=\rho_{r} \rho_{\theta}$. This ansatz yields two independent
systems of three equations each, one for the radial, and one for the angular dependences of the perturbations. The radial
equations become:

\begin{equation}
\frac{d \left( V_{r} /r^{1/2} \right)}{d r} = C_{r} r^{3/2} \frac{d \rho_{r}}{d r},
\end{equation}

\begin{equation}
\frac{d(r U_{r})}{d r} = C_{\theta} r^{2} \rho_{r},
\end{equation}

\begin{equation}
\frac{d \left( r^{1/2} V_{r} \right)}{d r} -\left( \frac{B C_{r}}{A}\right)\frac{d\left(r^{3/2} \rho_{r}\right)}{d r}=
C_{c} \frac{U_{r}}{r^{1/2}}.
\end{equation}

\noindent Where we have used the results of the angular ones:

\begin{equation}
V_{\theta} g(\theta) =\left( \frac{A}{C_{r}} \right) \rho_{\theta},
\end{equation}

\begin{equation}
\frac{d \rho_{\theta}}{d \theta} = \left( \frac{C_{\theta}}{A} \right) g(\theta) U_{\theta},
\end{equation}

\begin{equation}
\frac{d\left( sin(\theta) g(\theta) U_{\theta}\right)}{d \theta} =-C_{c} sin(\theta) g(\theta) V_{\theta}.
\end{equation} 

In splitting the radial and angular dependences of the perturbed continuity equation, eq.(24), we have used the result of
the angular equation of the perturbed radial Euler equation, eq.(30). The constants $C_{r}$, $C_{\theta}$ and $C_{c}$ are
the separation constants of the problem. Firstly we turn to the angular system, which allows an exact solution.

We can take eq.(32) and substitute into it the product $g(\theta)U_{\theta}$ from eq.(31), and the product $g(\theta)V_{\theta}$
from eq.(30) to obtain an equation involving only $\rho_{\theta}$:

\begin{equation}
\frac{d^{2} \rho_{\theta}}{d \theta^{2}} + cot(\theta)\frac{d \rho_{\theta}}{d \theta} + 
\left(\frac{C_{c}C_{\theta}}{C_{r}} \right)\rho_{\theta} =0,
\end{equation}

\noindent having solution:

\begin{equation}
\rho_{\theta} = c_{1} P_{m}(cos(\theta)) +c_{2} Q_{m}(cos(\theta)).
\end{equation}

In the above equation $c_{1}$ and $c_{2}$ are normalisation constants, and $P_{m}$ and $Q_{m}$ are the
Legendre polynomials of the first and second kinds, respectively. The index of these functions is
given by the relation $2m=(4C_{T}+1)^{1/2} -1$, where $C_{T}=(C_{c} C_{\theta}/C_{r})$. As typical of
separation of variables problems, we see the solution as an eigenvalue problem, with the separation
constants determining the order of the solution function. The requirement of axial symmetry forces $m$
to be even. Any such desired angular distribution for $\rho_{\theta}$ can now be constructed as an infinite
series of the above functions. For simplicity we analyse the case of $m=2$ $(C_{T}=6)$, $c_{1}=\bar{\rho}_{\theta}$, $c_{2}=0$:

\begin{equation}
\rho_{\theta} = \bar{\rho}_{\theta} (3 cos^{2}(\theta)-1 )
\end{equation}

\noindent This will result in slightly less material along the plane of
the disk, and slightly more along the poles, for the total density $\rho=\rho_{0}+\rho_{1}$, with respect to the background
state $\rho_{0}$. At this point eqs.(30) and (31) can be used to obtain:

\begin{equation}
V_{\theta} = \left( \frac{A \bar{\rho}_{\theta}}{C_{r}} \right) \left( \frac{3cos^{2}(\theta)-1}{g(\theta)}  \right)
\end{equation}

\begin{equation}
U_{\theta}=- \left( \frac{6 A \bar{\rho}_{\theta}}{C_{\theta}} \right) \left( \frac{cos(\theta)sin(\theta)}{g(\theta)}  \right)
\end{equation} 

We see the potential for jet formation already in the two above equations; if the background state is relatively devoid of material 
towards the poles, $g(\theta)$ could be small for $\theta=0$, and eq.(36) will lead to large radial velocities along the poles. Further,
eq.(37) shows that the angular velocity will always tend to zero both on the plane of the disk, and along the poles, where movement
will necessarily be radial, although not necessarily positive.

The case of the radial system is more cumbersome, as the linear operator which appears is of third order. However, 
in the Appendix we show through a first order stability analysis which terms can be safely ignored in the high-frequency 
approximation, and for small $r$, the region where jets develop. Equation (\ref{continuity1}) is used to show that 
the right-hand side of eq.(29) can be ignored to leading order, so that the resulting equation can be readily integrated to give:

\begin{equation}
V_{r}=\left( \frac{B C_{r}}{A} \right) r \rho_{r}.
\end{equation}

\noindent In the above we have taken the integration constant which appears as being equal to zero, from requiring
$V_{r}  \rightarrow 0$ for $\rho_{r} \rightarrow 0$. Substituting the above relation for $V_{r}$ into eq.(27) leads to:

\begin{equation}
\frac{d \rho_{r}}{d r} \left[ \left(\frac{B}{A} \right)r -r^{2} \right] +  \left(\frac{B}{A} \right)\frac{\rho_{r}}{2} =0.
\end{equation}

We see from equation (\ref{eulerv1}) in the Appendix that the second term on the left-hand side of the above equation can be 
dismissed, to leading order, in the regime of validity there established. We then obtain:

\begin{equation}
\rho_{r}=\bar{\rho}_{r} \left( \frac{\bar{r}_{\rho}}{r}\right)^{1/2},
\end{equation}

\begin{equation}
V_{r}=\left( \frac{B C_{r}}{A} \right) \bar{\rho}_{r} \bar{r}_{\rho}^{1/2} r^{1/2},
\end{equation}

\noindent where $\bar{r}_{\rho}$ is a characteristic radius at which $\rho_{r}=\bar{\rho}_{r}$. Now from equation (28), where 
all the terms are of the same order, as is shown in equation (\ref{euleru1}), we have

\begin{equation}
U_{r} = \left( \frac{2 C_{\theta} \bar{\rho}_{r} \bar{r}_{\rho}^{1/2} }{5} \right) r^{3/2}.
\end{equation}

In the above equation we have also taken the integration constant as zero, from requiring $U_{r} \rightarrow  0$ for
$r \rightarrow 0$. Choosing without loss of generality the two characteristic radii $\bar{r}$ and  $\bar{r}_{\rho}$ both
equal to $GM/c^{2}$, we can now write the full solution to the perturbation as:

\begin{equation}
\rho_{1}=\bar{\rho}_{J}\left( \frac{G M}{c^{2} r} \right)^{1/2} \left(3cos^{2}(\theta)-1 \right),
\end{equation}

\begin{equation}
V_{1}=c\left( \frac{\bar{\rho}_{J}}{\bar{\rho}_{0}} \right)\left( \frac{2c^{2}r}{GM} \right)^{1/2}
\left( \frac{3cos^{2}(\theta)-1 }{g_{\theta}} \right),
\end{equation}

\begin{equation}
U_{1}=-c\left( \frac{3 \bar{\rho}_{J}}{5\bar{\rho}_{0}} \right) \left( \frac{2c^{2}r}{GM} \right)^{3/2}
 \left( \frac{cos(\theta)sin(\theta)}{g_{\theta}} \right),
\end{equation}

\noindent where we have introduced $\bar{\rho}_{J}=\bar{\rho}_{\theta} \bar{\rho}_{r}$ and $g_{\theta}$ as the angular
part of $g(\theta)$, $g(\theta)/\bar{\rho}_{0}$.
The full solution to the perturbation can be seen to depend only on the two parameters $\bar{\rho}_{0}$ and $\bar{\rho}_{J}$,
normalisation constants for the densities of the background state and the perturbation, with the velocities
of the perturbation solution depending only, and linearly, on the ratio $Q=2^{1/2} \bar{\rho}_{J} / \bar{\rho}_{0}$,
which determines the validity regime of the perturbative approach through the condition $Q<2^{1/2}$.

As was already evident from eq.(37), we see form eq.(45) that the angular velocity will be zero only for $\theta=\pi/2$,  
$\theta=0$ and  $\theta=\pi$. Thus, movement along the plane of the disk will remain along the plane, but also, along 
the poles movement will be exclusively radial. This last point, together with the positive sign of the radial velocity 
along the poles (c.f. eq.(44) for $\theta=0$), provides for a well collimated jet along the poles. Hence, the background 
state proposed is seen to be unstable towards jet formation.

From eq.(44) we see that one has only to ask for a background state where matter density does not grow towards the poles, 
in order to obtain ejection velocities along them, which could become very large for flattened disks with relatively empty 
poles. The axial symmetry condition imposed on eq.(34) guarantees both axial symmetry and symmetry above and below the 
plane of the disk for the full solution. We see also that if one takes higher orders for $m$, the index of the Legendre 
polynomial solution to eq.(33), one obtains increasingly more critical angles at positions intermediary between $0$ and 
$\pi/2$ at which the angular velocity goes to zero. In fact, more complex geometries and asymmetric jets appear, as inferred 
observationally by e.g. Ferreira et al. (2006), if $m$ is taken as an arbitrary real number. However, modelling a situation 
where a polar jet dominates the ejection identifies $m=2$ as the leading order.

Notice that the qualitative behaviour of the solution is guaranteed by the exactness of the angular solution, the approximation
$r<2GM/c^{2}$ used for solving the radial problem will only introduce an error in the magnitudes of the velocities
outside of $r<2GM/c^{2}$, but will not change the fact that velocities will be of radial infall along the plane of the disk,
$\theta=\pi/2$, and of radial outflow along the poles, the jet solution for $\theta=0,\pi$. This is reinforced by the stability 
analysis shown in the appendix, where we show that the angular system of equations does not respond to a temporal perturbation
to the first order solution treated here, while the response of the radial system is merely the introduction of periodic and
bound factors multiplying the radial solutions given in this section.

The two constants of the problem, $\bar{\rho}_{0}$ and $\bar{\rho}_{J}$, can now be calculated once a choice of
$g(\theta)$ is specified, from the two conditions:

\begin{equation}
\dot{M}_{a} =2\pi \int_{0}^{\pi} \rho_{0}V_{0} sin(\theta) r^{2} d\theta,
\end{equation}

\begin{equation}
\dot{M}_{j} =2\pi \int_{0}^{\theta_{J}} \rho_{1}V_{1} sin(\theta) r^{2} d\theta,
\end{equation}

\noindent where $\theta_{J}$ is a suitable angle defining the opening of the jet, in all likelihood very small, as will
be see in the following section. In the above equations $\dot{M}_{a}$ gives the matter accretion rate onto the central
star, and $\dot{M}_{j}$ the matter ejection rate due to the jet. Dimensionally, the two quantities above will scale as:

\begin{equation}
\dot{M}_{a} =C_{a}  \frac{ \left( \pi G   M \right)^{2}  }{c^{3}} \bar{\rho}_{0}
\end{equation}

\begin{equation}
\dot{M}_{j} =C_{j}  \frac{ \left( \pi G   M \right)^{2}  }{c^{3}} \frac{\bar{\rho}_{J}^{2}}{\bar{\rho}_{0}},
\end{equation}

\noindent where $C_{a}$ and $C_{j}$ are two dimensionless constants which will depend on the choice of $g_{\theta}$,
and which would be expected to be of order 1.

Qualitatively, this type of model naturally furnishes a tight disk-jet connection (c.f. eq. 44) e.g., as now firmly established
in microquasars and AGN jets (see e.g. Marscher et al. 2002, Chatterjee et al. 2009). In the above systems bursts of enhanced jet 
activity are seen to follow temporal dips in disk luminosity output after small characteristic delay times. In the present
model, such a situation would be expected if the critical radius for transition to radial flow in the disk made a sudden 
transition to higher values. Again, the drop in disk output might not reflect the disk material disappearing (in this case
being swallowed by the central black hole, as sometimes proposed), but simply fading from view as heating mechanisms shut down,
then naturally enhancing jet activity as the effective $\dot{M}_{a}$ increases.


\section{Particular Solutions}

In order to present a sample of the trajectories expected in the model, we turn to the full solution to the
problem for $m=2$, eqs.(21) and (22), but written in dimensionless form:

\begin{equation}
\frac{d {\cal R} }{d {\cal T}} =\frac{-1}{ {\cal R}^{1/2}}+Q {\cal R}^{1/2} \left(\frac{3 cos^{2}(\theta)-1}{g_{\theta}} \right),
\end{equation}

\begin{equation}
\frac{d \theta}{d {\cal T}} =- \frac{6}{5}Q {\cal R}^{1/2} \left(\frac{sin(\theta)cos(\theta)}{g_{\theta}} \right).
\end{equation}

\noindent The above remain in spherical coordinates, where ${\cal R} = r c^{2}/GM$  and ${\cal T}=t c^{3}/GM$. Notice
that the threshold for jet development can now be stated as the condition $d{\cal R}/d{\cal T}>0$, 
$Q {\cal R} (3 cos^{2}(\theta)-1)> g_{\theta}$. A choice of $Q$ and $g_{\theta}$ now allows to numerically 
integrate trajectories. We take:

\begin{equation}
g_{\theta}=e^{-\left( \frac{\theta- \pi/2}{\sqrt{2} \theta_{0}}   \right)^{2}}
\end{equation}

\begin{figure}[!t]
\hskip -10pt \includegraphics[angle=0,scale=0.44]{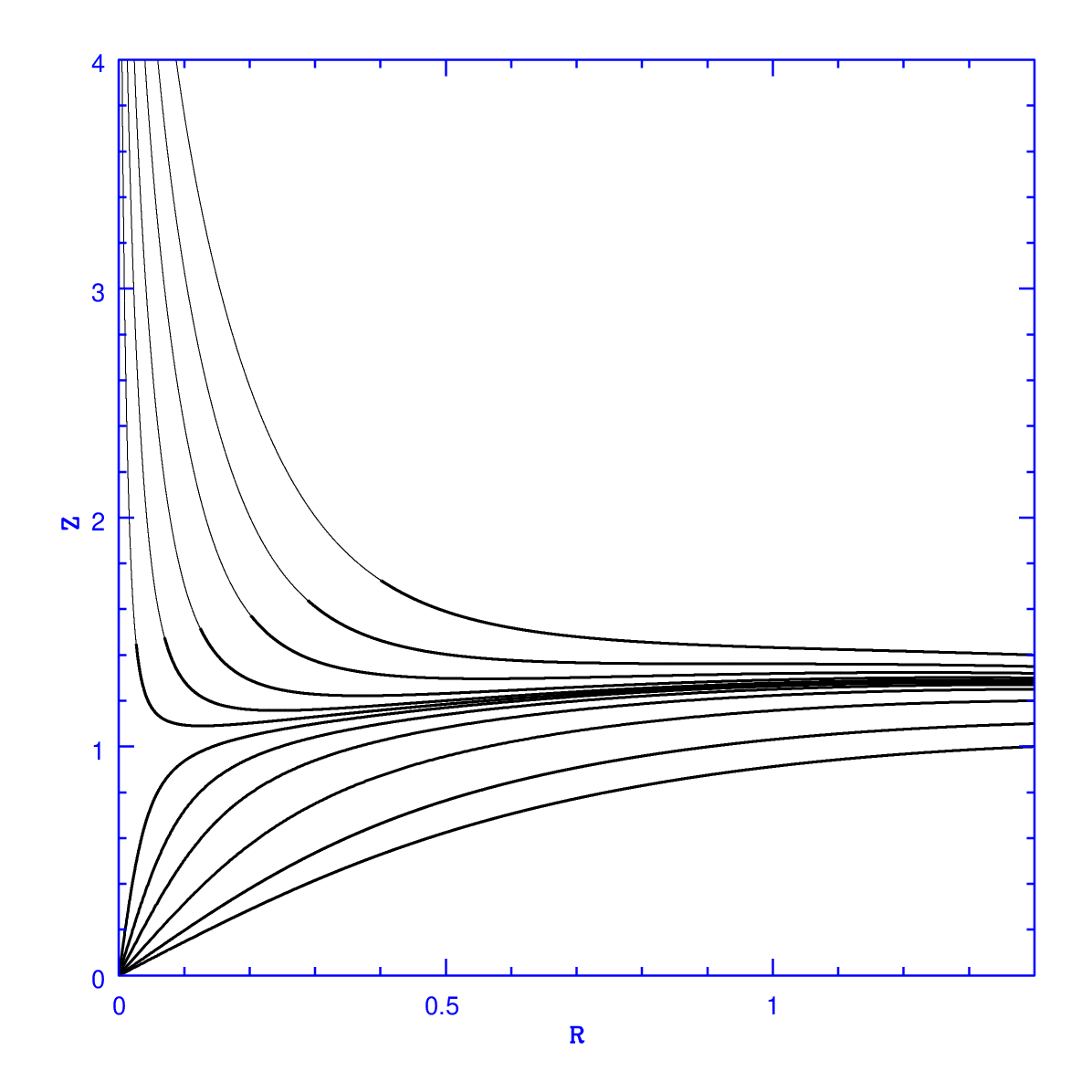}
\caption{The figure gives trajectories for the complete solution, for a range of initial vertical height positions at
$R=1.4$. The dimensionless parameters of the problem were chosen as $\theta_{0}=\pi/2$, $Q=0.3$. We see the lowermost six 
trajectories converging onto the central potential, but the six upper ones shooting up into a polar jet, with velocities
exceeding the local escape velocities.}
\end{figure}

\noindent as a model for the background state, where small values of $\theta_{0}$ will result in thin disks, while 
more spherical mass distributions correspond to large values of $\theta_{0}$. As a first example we take $Q=0.3$ 
($\bar{\rho}_{J} / \bar{\rho}_{0} =0.2$) and $\theta_{0}=\pi/2$, a fairly thick disk, with only very small changes in density from the
plane to the poles of a factor of 1.6, an almost spherically symmetric density configuration compatible with what appears in some
ADAF models. This value of $\theta_{0}$ is far from representing a flattened disk, and hence one which does not force the 
jet solution, see eq.(44). With these parameters, and initial conditions specified in dimensionless cylindrical coordinates 
as $R=1.4$ and $Z$ ranging from 1.0 to 1.4, we solve eqs.(50) and (51) through a finite differences scheme to plot figure 1. 

In spite of having taken $\bar{\rho}_{J} / \bar{\rho}_{0} <1$, the validity of the perturbative approach could be broken by 
the appearance of large velocities in the perturbation, breaking the validity condition $V_{1}/V_{0} <1$. We explicitly keep 
track of this condition, and show the appearance of perturbation radial velocities violating it with a change to thin lines 
in the figure, angular velocities for the perturbation will always remain of order $Q$, provided ${\cal R} <1$.

For this case, the two
lowermost curves present trajectories which all turn downwards to converge onto the central star. These are solutions which essentially 
follow the background state, infalling onto the bottom of the potential well. As one raises the initial value of $Z$ 
however, the following 4 curves begin to increasingly deviate from the free fall trajectories and, although still ending up 
at ${\cal R}=0$, clearly show a change in behaviour. A threshold is eventually crossed and curves of a very different type ensue, 
the six upper jet trajectories shown 
in figure 1. We see the pressure gradients associated with the distribution of matter in the background 
solution acting to break the fall of the incoming material, turn it back, and then accelerate it vertically 
through the vertical density gradients. These jet trajectories rapidly converge with height to eventually yield a well collimated 
structure.

Note that the region where the perturbation condition $V_{1}< V_{0}$ ceases to be valid appears only on the "jet" 
trajectories, and only after these have turned upwards, at points where in fact, velocities already exceed the local escape 
velocities. This point guarantees the appearance of a jet, even though the details of the flow in this jet region could differ 
somewhat from what is shown by the thin lines in the figure. The same applies to the $R<2$ validity region of the radial 
equation: the jet solution appears within the validity regime of all the approximations taken. the qualitative form of the 
full solution will not deviate much from what is shown in figure 1, due to the exact character of the angular solution.

\begin{figure}[!t]
\hskip -10pt \includegraphics[angle=0,scale=0.44]{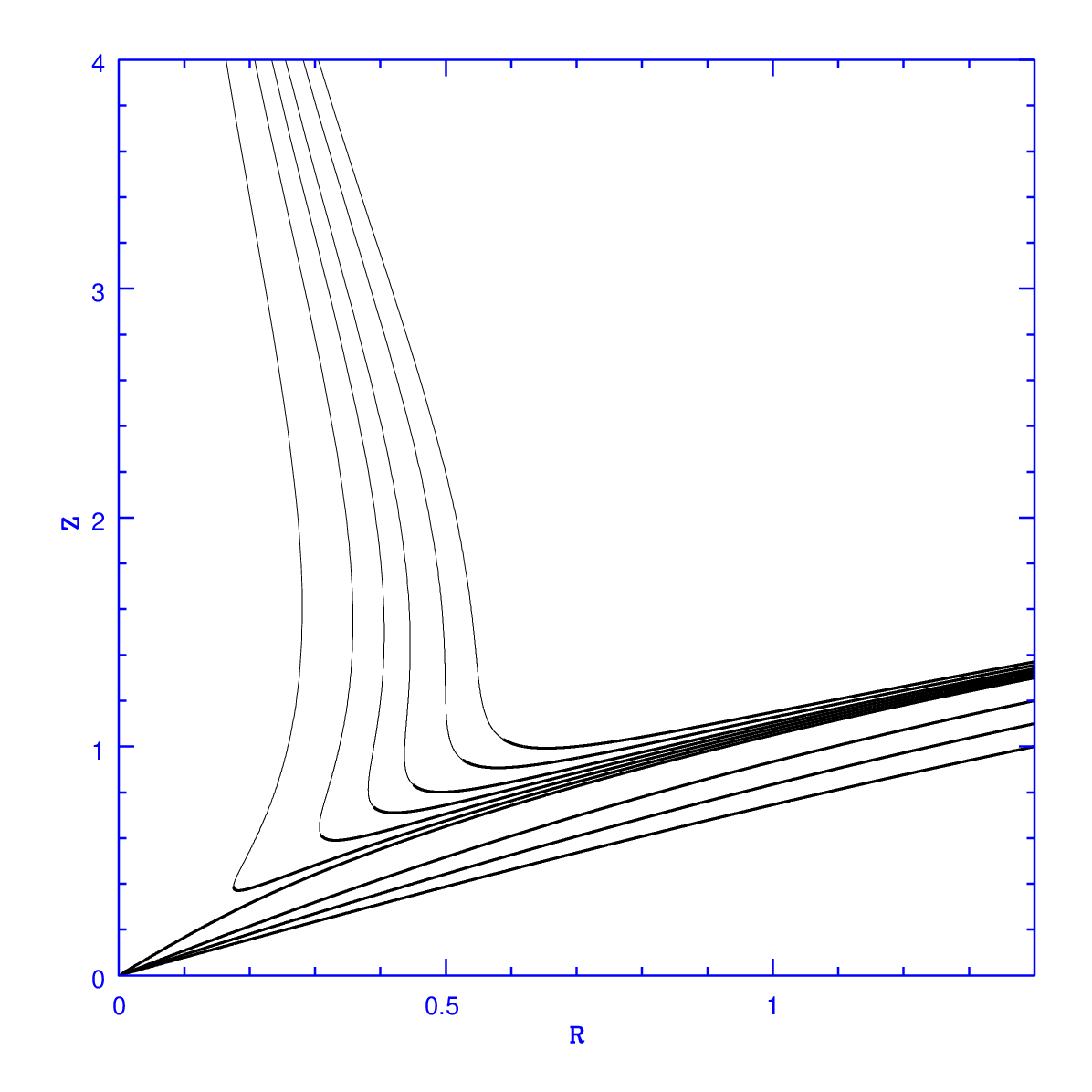}
\caption{The figure gives trajectories for the complete solution, for a range of initial vertical height positions at
$R=1.4$. The dimensionless parameters of the problem were chosen as $\theta_{0}=\pi/9$, $Q=0.01$. We see the lowermost four 
trajectories converging onto the central potential, but the six upper ones shooting up into a polar jet, with velocities
exceeding the local escape velocities.}
\end{figure}

Notice that the generic validity 
of the solution proposed does not require that all of the disk material should loose all of its angular momentum at the 
critical radius $R_{T}$, only that some of the disk material should loose most of its angular momentum at that point. Any
minor, residual angular momentum remaining on the disk fraction which forms the jet will only establish a finite jet 
cross-sectional area through the appearance of a centrifugal barrier, which will present only a small correction on the scenario
given here. Still, the empirical presence of spectroscopically studied accretion disks truncated interior to certain critical 
radii suggests the very substantial reduction of the shears in that region, as could happen if a transition to mostly radial 
flow takes place.

In going back to eqs.(50) and (51) we see that one expects

\begin{equation}
Q^{2} \sim \frac{\dot{M}_{j}}{\dot{M}_{a}}.
\end{equation}

\noindent It is reassuring that for the jets associated with T Tauri stars for example, values of $\dot{M}_{j} / \dot{M}_{a}$
of between 0.1 and 0.01 on average are observationally inferred and therefore of order $0.3<Q<0.1$ (e.g Hartigan et al. 1996, 
Gullbring et al. 1998, Hartigan et al. 2004, Ferreira et al. 2006). These are compatible with the value used to plot fig. (1), 
and hence ones which will readily yield jet solutions.


Figure 2 is analogous to fig. 1, but presents an example for a much more flattened disk having $\theta_{0}=\pi/9$,
a significantly flattened disk where the density contrast between plane and pole at fixed radius is now of a factor of 
2.5$\times 10^{4}$, orders of magnitude larger than in the last example. We take this time $Q=10^{-2}$. Again, we see 
the main features of the solution being well established in the region interior to $R<2$, with a jet solution again
established within the validity regime of all approximations used.

The long range stability and coherence of these structures lies outside
the scope of this work, and is in all probability furnished by a series of mechanisms extensively explored in the 
literature including angular momentum, magnetic fields and pressure containment of the surrounding medium, e.g. 
Begelman et al. (1984), Blandford (1990), Falle (1991), Kaiser \& Alexander (1997). 

In going to the more extreme jet phenomena associated with stellar black holes (e.g Mirabel \& Rodriguez 1999), 
quasars (e.g. Marscher 2002) and gamma ray bursts (e.g. M\'esz\'aros 2002), it is natural to expect the ideas presented 
here to apply, but amplified to a much more 
extreme regime by the appearance of corresponding relativistic and general relativistic effects, to first order, the 
shift in the divergence in the potential from $r=0$ in the Newtonian case to $r=r_{s}$ in the general relativistic one. 
It is therefore natural to expect purely hydrodynamic jet generation mechanism to apply across all classes of astrophysical 
objects, specially given the qualitative scalings and similarities which appear over all astrophysical jet classes 
(e.g. Miarabel \& Rodriguez 1999, Mendoza, Hernandez \& Lee 2005), in addition to the magnetically driven processes 
traditionally found in the literature. 

Notice from eq.(44) the intimate link between the jet velocity and the physical state of the infalling material, $c$
and $\dot{M}_{a}$ through  $\bar{\rho}_{0}$. This implies that temporal variations in the parameters of the infalling material
will result in temporal variations in the density and velocity of the jet material, in a way described by eqs.(43) and (44).
The above can serve as a physical description of the key processes relevant to the formation of internal shocks in 
astrophysical jets, the main ingredient behind phenomena such as HH objects and gamma ray bursts.

Although the solution to the problem presented is rigorous, and the appearance of the resulting hydrodynamical jets solid, 
the relevance of our model is clearly constrained by the validity, or otherwise, of the starting hypothesis; most critically,
the appearance of an inner region in free fall. Although plausibility arguments for the appearance of this regime have been 
given, the reality of it is certainly not assured. Still, given the absence of a fully satisfactory and general solution to 
the problem of jet generation, and given the clear and transparent physics of the one shown here, we believe it is interesting
to present such a simple hydrodynamical jet solution.

\section{Conclusions}

We show that given a radially infalling accretion, a purely hydrodynamical
jet ensues. 

We calculate the condition for the transition from quasi-circular to quasi-radial
flow in a standard accretion disk, and show it will always occur for power law surface density profiles
of the form $\Sigma \propto R^{-n} $, interior to a critical radius, provided $n>1/2$.

Comparison with inferred inner holes in observed accretion disks yields results consistent with 
our estimates for the above transition radius, the point where shears in the flow would be
substantially reduced.

Well collimated jets readily appear, proving the existence of purely hydrodynamical mechanisms for the
generation of astrophysical jets.

\section{Acknowledgments}
Xavier Hern\'andez acknowledges the hospitality of the Observatoire de Paris for the duration of a
sabbatical stay during which many of the ideas presented here were first developed, and financial
support from project PAPIIT IN103011 DGAPA UNAM. Pablo L. Rend\'on
and Rosa Rodr\'{\i}guez acknowledge financial support from project  PAPIIT IN110411 DGAPA UNAM. Antonio Capella 
acknowledges financial support from project PAPIIT IN101410 DGAPA UNAM.

\appendix
\section{Appendix: Stability Analysis}

We study the stability of the system by means of a standard linear perturbation approach. Furthermore, we develop here a 
first order analysis that also serves to justify the approximations made in section \ref{hydro} of the main 
body of the paper. We follow the approach used by Papaloizou \& Pringle (1984) to study the stability of this system by 
means of linear perturbation of the time-dependent equations of the fluid, as opposed to the steady-state versions stated 
previously as equations (24)-(26). The time-dependence of the perturbed quantities is written as $\exp (i \omega t)$, 
where $\omega$ is a temporal frequency. We then obtain

\begin{eqnarray}
i \omega  \rho_1 \, \frac{r^2}{\bar{r}^{3/2}} & + & g(\theta) \frac{\partial \left(r^{1/2} V_{1}\right)}{\partial r} -  
B\frac{\partial \left( r^{3/2} \rho_{1} \right)}{\partial r}
\nonumber \\
& = & \frac{-1}{r^{1/2} sin(\theta)} \frac{\partial \left( sin(\theta) g(\theta) U_{1} \right) }{\partial \theta},
\label{continuity}
\end{eqnarray}

\noindent
for the continuity equation, and

\begin{equation}
i \omega \left( \frac{1}{2GM} \right)^{1/2} V_1 + \frac{\partial \left( V_{1}/r^{1/2} \right) }{\partial r} = 
\frac{A r^{3/2}}{g(\theta)} \frac{\partial \rho_{1}}{\partial r},
\label{eulerv}
\end{equation}

\begin{equation}
-i \omega r \left( \frac{1}{2GM} \right)^{1/2} U_1 + \frac{\partial \left( r U_{1}  \right)}{\partial r} = 
\frac{A r^{2}}{g(\theta)} \frac{\partial \rho_{1}}{\partial \theta},
\label{euleru}
\end{equation}

\noindent
for the radial and angular components of the Euler equation, respectively. We now proceed as before, separating variables 
in such a way that we require the angular components $\rho_{\theta}$, $V_{\theta}$ and $U_{\theta}$ to be dimensionless, while 
the radial components $\rho_r$, $V_r$ and $U_r$ have the same dimensions as the original functions. Encouragingly, the 
equations for the angular components are exactly the same as those obtained for the steady-state analysis, equations (30)-(32). 
We conclude that the angular components are not time-dependent and are thus unconditionally stable, reinforcing the overall 
character of the solution presented in section 3. The resulting equations 
for the radial components do include an extra term due to the temporal dependence. In order to establish the relative magnitude 
of the different terms in these equations, we shall rewrite them in dimensionless form, using the following scalings for that effect:

\begin{equation}
r^*=\frac{r}{r_0}, \quad \rho^*=\frac{\rho_r}{\rho_0}, \quad V^*=\frac{V_r}{\omega \bar{r}}, \quad U^*=\frac{U_r}{\omega r_0},
\label{scaling}
\end{equation}

\noindent
where $\rho_0$ is a characteristic density, $\bar{r}=2GM/c^2$, and $r_0=(2GM/\omega^2)^{1/3}$. We now consider the high-frequency 
limit, as Papaloizou \& Pringle (1984) have also done. Notice then that $\epsilon = r_0/\bar{r} = c^2/(2GM \omega)^{2/3} \ll 1$. 
Equation (\ref{continuity}) may now be rewritten as

\begin{equation}
\frac{\partial ({r^*}^{1/2} V^*)}{\partial r^*} - \beta \frac{\partial ({r^*}^{3/2} \rho^*)}{\partial r^*} + i \beta {r^*}^{2} \rho^* 
= \epsilon C_c \frac{U^*}{{r^*}^{1/2}},
\label{continuity1} 
\end{equation}

\noindent
where $\beta = C_r \rho_0 (2GM/\omega^5)^{1/3}$. If we now require $C_c \sim 1$ and $C_r \sim \rho_0^{-1} (\omega^5/2GM)^{1/3}$, 
we observe that all the terms on the left-hand side of the equation are $O(1)$, whereas the term on the right-hand side is 
$O(\epsilon)$. Thus, at leading order the term on the right-hand side vanishes.

Using this same scaling, equation (\ref{eulerv}) is now

\begin{equation}
\frac{\partial ({r^*}^{-1/2} V^*)}{\partial r^*} + iV^* = \epsilon \beta {r^*}^{3/2} \frac{\partial \rho^*}{\partial r^*},
\label{eulerv1}
\end{equation}

\noindent
where we observe that the term on the right-hand side is $O(\epsilon)$. Again, at $O(1)$, the right-hand side of the 
equation vanishes, and we may solve the equation to obtain

\begin{equation}
V^*= \alpha_1 {r^*}^{1/2} \exp \left( -i \frac{2}{3} {r^*}^{3/2} \right),
\label{vsolution}
\end{equation}

\noindent
where $\alpha_1$ is a constant. Substituting the real part of solution (\ref{vsolution}) in (\ref{continuity1}) we 
now solve for $\rho^*$,

\begin{eqnarray}
\rho^* & = & \frac{\alpha_1}{2 \beta {r^*}^{1/2}} \left[  \exp \left( i \frac{2}{3} {r^*}^{3/2} \right) + \frac{1}{2} 
\exp \left( -i \frac{2}{3} {r^*}^{3/2} \right)\right]
\nonumber \\
& + & \alpha_2 {r^*}^{-3/2} \exp \left( -i \frac{2}{3} {r^*}^{3/2} \right) + \frac{i \alpha_1 r^*}{5 \beta} \exp \left( i 
\frac{2}{3} {r^*}^{3/2} \right)
\nonumber \\
& + & \frac{i^{4/3} \alpha_1}{4 \cdot 6^{1/3} \, \beta {r^*}^{3/2}} \exp \left( i \frac{2}{3} {r^*}^{3/2} \right) \Gamma 
\left( \frac{2}{3}, i \frac{4}{3} {r^*}^{3/2} \right),
\label{rhosolution}
\end{eqnarray}

\noindent
where $\alpha_2$ is a constant, and $\Gamma$ is the incomplete gamma function. The real parts of these solutions are 
the dimensionless forms of the solutions we obtained for $V_r$ and $\rho_r$ in section \ref{hydro} of this paper, 
modulated by an oscillating and bounded function.

Finally, in dimensionless form, equation (\ref{euleru}) is 

\begin{equation}
\frac{\partial (r^* U^*)}{\partial r^*} -i {r^*}^{3/2} U^* = \beta' {r^*}^{2} \rho^*.
\label{euleru1}
\end{equation}

\noindent
where $\beta' = C_{\theta} \rho_0 (2GM/\omega^5)^{1/3}$. If we now require $C_{\theta} \sim \rho_0^{-1} (\omega^5/2GM)^{1/3}$, 
all three terms are $O(1)$. Notice that the orders of magnitude chosen here for the separation constants give as a result 
$C_T \sim 1$, so that the choice of value given to $C_T$ in section \ref{hydro} is consistent with our analysis. Taking 
the real part of the solution given in (\ref{rhosolution}) and substituting in the above equation, we may solve to obtain


\begin{eqnarray}
U^* & = & \exp \left( i\frac{2}{3} {r^*}^{3/2} \right) \left[ \frac{\alpha_3}{r^*}  + \frac{1}{10} \frac{\alpha_1 
\beta'}{\beta} {r^*}^{3/2} \right] 
\nonumber \\
& + & \frac{\alpha_1 \beta'}{\beta} \frac{i^{1/3}}{8 \cdot 6^{1/3} r^*} \exp \left( i \frac{2}{3} {r^*}^{3/2} \right) 
\Gamma \left( \frac{2}{3}, i \frac{4}{3} {r^*}^{3/2} \right)
\nonumber \\
& + &   \frac{\alpha_1 \beta'}{\beta} \frac{i}{8} \exp \left( -i \frac{2}{3} {r^*}^{3/2} \right),  
\end{eqnarray}

\noindent
where $\alpha_3$ is a constant. Considering the real part of the above solution, we again find that we have the 
dimensionless form of the solution obtained for $U_r$ earlier in this paper, modulated by an oscillating, bounded function. 

Thus, the effect of linear perturbations on the time-dependent system is to provoke bounded oscillations around the 
steady-state solutions for density and both angular and radial velocities. We conclude that the system is neutrally stable.

Our choice of scalings has permitted us to write the equations of the fluid in dimensionless form in terms of only
 three dimensionless parameters, $\epsilon$, $\beta$ and $\beta'$. In this form, as is expected, it becomes rather 
more straightforward to establish the leading-order terms in the equations. The hypothesis of small radial distance 
translates into the smallness of $\epsilon$, and $\beta$ and $\beta'$, formally very similar, are related to the 
separation constants. It should be pointed out that the dimensional constants $C_r$ and $C_\theta$ are independently
 required to be of the same order, and by simply letting the dimensionless constant $C_c$ be $O(1)$, the sole 
restriction placed upon these three constants in section \ref{hydro}, that $C_T$ also be $O(1)$, is fulfilled.
 We can then consistently fix the order of the Legendre polynomial solutions presented in section \ref{hydro} as 
a function of the particular values given to the separation constants, a behaviour typically associated with the 
method of separation of variables.

\end{document}